\documentclass[aps, pre, superscriptaddress, onecolumn]{revtex4}
\usepackage{amsmath}
\usepackage[utf8]{inputenc}
\usepackage{bm}
\usepackage{amsfonts}
\usepackage{amsthm}
\usepackage{xcolor}
\usepackage{soul}
\usepackage[breaklinks=true,colorlinks=true,linkcolor=blue,urlcolor=blue,citecolor=blue]{hyperref}

\newcommand{\defeq}{\mathrel{\mathop:}=}
\renewcommand{\eqref}[1]{Eq.~(\ref{#1})}

\newcommand{\K}{\mathcal{K}}
\renewcommand{\H}{\mathcal{H}}

\newcommand{\vth}{v_{\text{th}}}
\DeclareMathOperator*{\argmax}{argmax}

\begin{document}

\title{Kappa distribution from particle correlations\\ in non-equilibrium, steady-state plasmas}

\author{Sergio Davis}
\affiliation{Research Center in the intersection of Plasma Physics, Matter and Complexity (P$^2$mc),\\ Comisi\'on Chilena de Energía Nuclear, Casilla 188-D, Santiago, Chile}
\affiliation{Departamento de F\'isica, Facultad de Ciencias Exactas,\\ Universidad Andres Bello, Sazi\'e 2212, piso 7, 8370136, Santiago, Chile.}
\email{sergio.davis@cchen.cl}

\author{Gonzalo Avaria}
\affiliation{Departamento de F\'isica, Universidad Técnica Federico Santa Mar\'ia, Av. Vicuña Mackenna 3939, Santiago, 8940000, Chile}

\author{Biswajit Bora}
\affiliation{Research Center in the intersection of Plasma Physics, Matter and Complexity (P$^2$mc),\\ Comisi\'on Chilena de Energía Nuclear, Casilla 188-D, Santiago, Chile}
\affiliation{Departamento de F\'isica, Facultad de Ciencias Exactas,\\ Universidad Andres Bello, Sazi\'e 2212, piso 7, 8370136, Santiago, Chile.}

\author{Jalaj Jain}
\affiliation{Research Center in the intersection of Plasma Physics, Matter and Complexity (P$^2$mc),\\ Comisi\'on Chilena de Energía Nuclear, Casilla 188-D, Santiago, Chile}
\author{Jos\'e Moreno}
\author{Cristian Pavez}
\author{Leopoldo Soto}
\affiliation{Research Center in the intersection of Plasma Physics, Matter and Complexity (P$^2$mc),\\ Comisi\'on Chilena de Energía Nuclear, Casilla 188-D, Santiago, Chile}
\affiliation{Departamento de F\'isica, Facultad de Ciencias Exactas,\\ Universidad Andres Bello, Sazi\'e 2212, piso 7, 8370136, Santiago, Chile.}

\date{\today}

\begin{abstract}
Kappa-distributed velocities in plasmas are common in a wide variety of settings, from low-density to high-density plasmas. To date, they have been found mainly in 
space plasmas, but are recently being considered also in the modelling of laboratory plasmas. Despite being routinely employed, the origin of the kappa distribution 
remains, to this day, unclear. For instance, deviations from the Maxwell-Boltzmann distribution are sometimes regarded as a signature of the non-additivity of the 
thermodynamic entropy, although there are alternative frameworks such as superstatistics where such an assumption is not needed. In this work we recover the kappa 
distribution for particle velocities from the formalism of non-equilibrium steady-states, assuming only a single requirement on the dependence between the kinetic 
energy of a test particle and that of its immediate environment. Our results go beyond the standard derivation based on superstatistics, as we do not require any 
assumption about the existence of temperature or its statistical distribution, instead obtaining them from the requirement on kinetic energies. All of this suggests 
that this family of distributions may be more common than usually assumed, widening its domain of application in particular to the description of plasmas from fusion 
experiments. Furthermore, we show that a description of kappa-distributed plasma is simpler in terms of features of the superstatistical inverse temperature 
distribution rather than the traditional parameters $\kappa$ and the thermal velocity $\vth$.
\end{abstract}

\maketitle

\section{Introduction}

Modelling the velocity distribution of particles in a non-equilibrium, steady-state plasma is an interesting challenge from both theoretical and practical points 
of view~\cite{Hellberg2002,Appelbe2011,Milder2019,Zorondo2022}. In general, particles in a steady state plasma do not follow the typical Maxwell-Boltzmann distribution 
of velocities one would expect in an equilibrium system but, instead, their velocities are described by more general families of distributions. Among them, the kappa 
distribution, a power-law distribution with long tails describing highly energetic particles, appears predominantly in space plasmas~\cite{Pierrard2010,Livadiotis2017}. 
These are weakly collisional plasmas in steady states found in the Earth's magnetosphere~\cite{Antonova2008,Espinoza2018,Kirpichev2020,Eyelade2021} and plasma sheet~\cite{Christon1988,Stepanova2015,Espinoza2018}, as well as in the solar wind~\cite{Maksimovic1997b,Nicolaou2019,ZentenoQuinteros2021}. Beyond Earth, kappa distributions have also been found in other planetary atmospheres~\cite{Baluku2011,Carbary2014,Nicolaou2014,Kandpal2018} and the interstellar medium~\cite{Raymond2010,Nicholls2017}.

\nocite{Oylukan2023}

From the point of view of laboratory plasmas, even when the energy distribution of suprathermal ions in fusion plasmas, such as the ones generated by Z-pinch 
discharges, has been known for decades to be well-described by power laws~\cite{Potter1971,Bernstein1972,Stygar1982,Vikhrev2007,Vikhrev2012}, only recently 
kappa distributions have being proposed in this context as possible statistical models~\cite{Knapp2013,Klir2015}.

\vspace{5pt}
\noindent
For the velocity $\bm v$ of a particle of mass $m$, the kappa distribution is commonly written in the form
\begin{equation}
\label{eq:kappa}
P(\bm v|\kappa, \vth) = \frac{1}{\eta_\kappa(\vth)}\left[1+\frac{1}{\kappa-\frac{3}{2}}\frac{\bm{v}^2}{\vth^2}\right]^{-(\kappa+1)}
\end{equation}
where $\kappa \geq 0$ is a shape parameter, sometimes referred to as the \emph{spectral index}, $\vth$ is the thermal velocity~\cite{Bellan2006},
\begin{equation}
\label{eq:vth}
\vth \defeq \sqrt{\frac{2 k_B T}{m}},
\end{equation}
and $\eta_\kappa(\vth)$ is a normalization constant given by
\begin{equation}
\label{eq:etakappa}
\eta_\kappa(\vth) \defeq \left(\sqrt{\pi(\kappa-3/2)}\vth\right)^3 \frac{\Gamma(\kappa-1/2)}{\Gamma(\kappa+1)}.
\end{equation}

\noindent
In the limit $\kappa \rightarrow \infty$, the kappa distribution in \eqref{eq:kappa} reduces to the Maxwell-Boltzmann distribution,
\begin{equation}
P(\bm v|m, T) = \left(\sqrt{\frac{m}{2\pi k_B T}}\right)^3\exp\Big(-\frac{m\bm{v}^2}{2 k_B T}\Big),
\end{equation}
precisely the distribution expected in equilibrium at temperature $T$. However, for finite $\kappa$ the interpretation of the parameter $T$ in
\eqref{eq:vth} is not straightforward~\cite{Nicolaou2016,Sattin2018}, mainly because there are multiple admissible definitions of temperature and not all of them agree 
with $T$.

Although the presence of kappa distributions in plasmas has been traditionally explained~\cite{Leubner2002, Leubner2004} by the use of non-extensive statistical 
mechanics, also known as Tsallis statistics~\cite{Tsallis2009c}, more recent frameworks such as superstatistics~\cite{Beck2003,Beck2004} can recover them in a direct 
manner. Moreover, recently we have shown~\cite{Davis2019b} that superstatistics arises as a natural description for collisionless plasmas in non-equilibrium steady 
states, providing support to recent efforts~\cite{Ourabah2015,Ourabah2020b,Sanchez2021} in establishing a foundational basis for steady-state distributions in plasmas using 
superstatistics as a starting point.

Despite these advances, superstatistics still requires the assumption of a gamma distribution for the inverse temperature $\beta \defeq 1/(k_B T)$ in
order to recover Tsallis statistics, and in particular, the kappa distributions. This particular choice of temperature distribution is referred to as $\chi^2$-superstatistics. Several mechanisms aiming to explain the origin of this $\chi^2$ family of superstatistics have been proposed in the literature since the theory was 
originally presented. For instance, by using the fact that a sum of squares of Gaussian random variables becomes gamma distributed as the number of such variables grows 
large~\cite{Beck2005}, or by invoking Jaynes' maximum entropy principle on the distribution $P(\beta|S)$ under macroscopic constraints~\cite{VanDerStraeten2008}.

Motivated by this somewhat unsatisfactory assumption of gamma-distributed inverse temperatures, in this work we delve deeper into the formalism established in 
Ref.~\cite{Davis2019b}, by connecting it with more recent theoretical developments~\cite{Davis2020,Davis2022b} on the structure of superstatistics. 
In particular, we show that the assumption of a gamma distribution for $\beta$ can be replaced by a simpler, and perhaps more fundamental, assumption on the 
dependence between the kinetic energy of a test particle and that of its surrounding environment.

In the following section we provide a brief account of the superstatistical formalism and we connect it with a generalized definition of temperature for steady 
states~\cite{Davis2023b}, namely the \emph{fundamental inverse temperature} function $\beta_F$.

\nocite{Davis2019} 

\section{Non-equilibrium steady states and superstatistics}
\label{sec:super}

Steady states are a special kind of non-equilibrium states which are time-independent, that is, where the non-equilibrium probability density of microstates 
$p(\bm{\Gamma}; t)$ at a time $t$ reduces to $p(\bm{\Gamma})$. In particular, we will consider steady states where $p(\bm{\Gamma})$ depends on $\bm{\Gamma}$ only 
through the Hamiltonian $\H(\bm \Gamma)$, and we will write their probability density as
\begin{equation}
\label{eq:rho}
P(\bm \Gamma|S) = \rho(\H(\bm \Gamma)),
\end{equation}
where $\rho$ is the \emph{ensemble function}, and $S$ denotes the set of parameters that uniquely define the steady state. 

Within this general framework, superstatistics is a natural extension of statistical mechanics to steady states in the form given by \eqref{eq:rho}.
Besides nonequilibrium plasmas, it has been successfully used in high-energy physics~\cite{Jizba2010,Ayala2018}, anomalous diffusion~\cite{Itto2014,Metzler2020}, 
cosmology and gravitation~\cite{Ourabah2019}, turbulence~\cite{Reynolds2003,VanDerStraeten2009,Gravanis2021}, seismicity~\cite{Iliopoulos2019}, bioinformatics~\cite{Costa2022}, as well as phenomena of interest in engineering such as the electrical fluctuations of power grids~\cite{Schafer2018}.

\noindent
In superstatistics, the canonical ensemble
\begin{equation}
P(\bm \Gamma|\beta) = \frac{\exp(-\beta \H(\bm \Gamma))}{Z(\beta)}
\end{equation}
is replaced by a superposition of canonical ensembles at different temperatures. The inverse temperature $\beta$ is promoted
from a constant to a random variable with probability density $P(\beta|S)$, such that its joint distribution with the microstates is given by
\begin{equation}
\label{eq:super}
P(\bm \Gamma, \beta|S) = P(\bm \Gamma|\beta)P(\beta|S) = \left[\frac{\exp\big(-\beta \H(\bm \Gamma)\big)}{Z(\beta)}\right] P(\beta|S).
\end{equation}

\noindent
By marginalization of $\beta$, the distribution of microstates becomes
\begin{equation}
\label{eq:super_margin}
P(\bm \Gamma|S) = \int_0^\infty d\beta P(\beta|S)\left[\frac{\exp(-\beta \H(\bm \Gamma))}{Z(\beta)}\right],
\end{equation}
which has the form of \eqref{eq:rho} with an ensemble function
\begin{equation}
\label{eq:laplace}
\rho(E) = \int_0^\infty d\beta f(\beta)\exp(-\beta E),
\end{equation}
that is the Laplace transform of the \emph{superstatistical weight function} $f(\beta)$, defined by
\begin{equation}
\label{eq:supweight}
f(\beta) \defeq \frac{P(\beta|S)}{Z(\beta)}.
\end{equation}

The distinction between $f(\beta)$ and $P(\beta|S)$ is an important one. The formulation of superstatistics as in \eqref{eq:super_margin} is known as 
type-B superstatistics, and is the standard version in use nowadays~\cite{Beck2004}. The original formulation~\cite{Beck2003}, now known as type-A superstatistics, 
defines $\rho$ as in \eqref{eq:laplace} but $f(\beta)$ itself is taken as the probability density for $\beta$. This led to inconsistencies with the application 
of the sum and product rule of probability~\cite{Sattin2006}.

Among the possible families of distributions compatible with superstatistics, three \emph{universality classes} have been shown to be especially relevant for 
non-equilibrium systems: the so-called $\chi^2$-superstatistics where $f(\beta)$ has the form of a gamma distribution, log-normal superstatistics and inverse-gamma superstatistics. Arguably the most predominant case is the $\chi^2$-superstatistics, as it leads to the $q$-canonical ensemble of Tsallis statistics, and in particular 
to the kappa distribution. However, the log-normal superstatistics has been found in the context of turbulence~\cite{Reynolds2003,Jung2005}, and in stellar systems~\cite{Ourabah2022}, among several other contexts. On the other hand, the inverse-gamma superstatistics has been successfully employed to described the 
thermodynamics of small molecules~\cite{Dixit2013} and the dynamics of protein diffusion~\cite{Itto2021}.

Using the definition in \eqref{eq:laplace} we can write $\rho(E) = \mathcal{L}\big\{f\big\}(E)$ and, conversely, $f(\beta) = \mathcal{L}^{-1}\big\{\rho\big\}(\beta)$. 
An important consequence of this is that $\rho$ is completely determined by $f$ and viceversa, and as the latter depends on both the inverse temperature distribution 
and the partition function, then both aspects together define the form of the statistical ensemble $P(\bm \Gamma|S)$.

Let us now consider a composite system, divided into subsystems $A$ and $B$ such that $\bm{\Gamma} = (\bm{\Gamma}_A, \bm{\Gamma}_B)$, and where the Hamiltonian
of the entire system is of the form
\begin{equation}
\H(\bm{\Gamma}_A, \bm{\Gamma}_B) = \H_A(\bm{\Gamma}_A) + \H_B(\bm{\Gamma}_B).
\end{equation}

Please note that, because we are considering a superstatistical ensemble function $\rho(E)$ of the form in \eqref{eq:super_margin}, it is no longer true that additive 
subsystems have a joint distribution that is the product of their marginal distributions. That is, in general,
\begin{equation}
\rho(E|S) \neq \rho(E_A|S)\rho(E_B|S).
\end{equation}

\noindent
The statistical independence of subsystems only remains true for the canonical ensemble, where \[\rho(E_A+E_B|\beta_0) = \rho(E_A|\beta_0)\rho(E_B|\beta_0).\]

However, it is easy to show in the general case that $P(\beta|S)$ is a universal property of the entire system and its parts, that is, the same $P(\beta|S)$ function is involved 
when expressing the ensemble function of an arbitrary subsystem $\nu$ as in \eqref{eq:super_margin},
\begin{equation}
\label{eq:super_margin_alpha}
P(\bm{\Gamma}_\nu|S) = \int_0^\infty d\beta P(\beta|S)\left[\frac{\exp(-\beta \H_\nu(\bm{\Gamma}_\nu))}{Z_\nu(\beta)}\right],
\end{equation}
noting that $P(\beta|S)$, unlike $\H_\nu$ and $Z_\nu$, does not carry the subindex $\nu$. It follows that $f_\nu(\beta)$, being the ratio between $P(\beta|S)$ and 
the partition function $Z_\nu(\beta)$, will in fact be dependent on the details of the subsystem. We can show that \eqref{eq:super_margin_alpha} holds as follows. Let the 
composite system be described by an inverse temperature distribution $P(\beta|S)$. Then we have
\begin{equation}
\label{eq:super_AB}
P(\bm{\Gamma}_A, \bm{\Gamma}_B|S) = \int_0^\infty d\beta P(\beta|S)\left[\frac{\exp(-\beta (\H_A+\H_B))}{Z_{AB}(\beta)}\right],
\end{equation}
and the marginal distribution of $\bm{\Gamma}_A$ is given by
\begin{equation}
\begin{split}
P(\bm{\Gamma}_A|S) & = \int d\bm{\Gamma}_B\int_0^\infty d\beta P(\beta|S)\left[\frac{\exp(-\beta (\H_A+\H_B))}{Z_{AB}(\beta)}\right] \\
& = \int_0^\infty d\beta P(\beta|S)\exp\big(-\beta \H_A(\bm{\Gamma}_A)\big)\int d\bm{\Gamma}_B\left[\frac{\exp\big(-\beta \H_B(\bm{\Gamma}_B)\big)}{Z_{AB}(\beta)}\right] \\
& = \int_0^\infty d\beta P(\beta|S)\frac{Z_B(\beta)}{Z_{AB}(\beta)}\exp\big(-\beta \H_A(\bm{\Gamma}_A\big)),
\end{split}
\end{equation}
that is,
\begin{equation}
\label{eq:super_A}
P(\bm{\Gamma}_A|S) = \int_0^\infty d\beta P(\beta|S)\left[\frac{\exp(-\beta \H_A(\bm{\Gamma}_A))}{Z_A(\beta)}\right],
\end{equation}
where we have used the well-known factorization of the partition function for additive systems, $Z_{AB}(\beta) = Z_A(\beta)Z_B(\beta)$. We see from \eqref{eq:super_A} and
\eqref{eq:super_AB} that the subsystem $\bm{\Gamma}_A$ is governed by the same inverse temperature distribution $P(\beta|S)$ as the composite system $(\bm{\Gamma}_A, \bm{\Gamma}_B)$ and, 
because the choice of $A$ and $B$ is arbitrary, it follows that any possible subsystem is governed by the same $P(\beta|S)$. In the following we will use this fact to recover 
subsystem-independent parameters for the kappa distribution describing the velocity of a single particle.

We will now define the \emph{fundamental inverse temperature} function $\beta_F$, motivated by the conditional distribution of $\beta$ given a fixed energy $E$. First, note that 
the distribution of energy in an steady state given by \eqref{eq:rho} is
\begin{equation}
\label{eq:edist_rho}
P(E|S) = \Big<\delta(E-\H)\Big>_S = \int d\bm{\Gamma}\rho(\H(\bm \Gamma))\delta(E-\H(\bm \Gamma)) = \rho(E)\Omega(E),
\end{equation}
where $\Omega(E) \defeq \int d\bm{\Gamma}\delta(E-\H(\bm \Gamma))$ is the density of states associated to $\H$. Now, from Bayes' theorem~\cite{Jaynes2003,Sivia2006} we obtain
\begin{equation}
\label{eq:pbeta_bayes}
P(\beta|E, S) = \frac{P(\beta|S)P(E|\beta, S)}{P(E|S)}
\end{equation}
and, because exact knowledge of $\beta$ supersedes the state of knowledge $S$, we can replace $P(E|\beta, S)$ in the numerator with the usual canonical distribution of energy,
\begin{equation}
\label{eq:edist_canon}
P(E|\beta) = \frac{\exp(-\beta E)\Omega(E)}{Z(\beta)},
\end{equation}
a particular case of \eqref{eq:edist_rho} with $\rho(E) = \exp(-\beta E)/Z(\beta)$. Therefore, replacing \eqref{eq:edist_canon} and \eqref{eq:edist_rho} into 
\eqref{eq:pbeta_bayes} and cancelling the factor $\Omega(E)$, we have
\begin{equation}
\label{eq:pbeta_cond}
P(\beta|E, S) = \frac{f(\beta)\exp(-\beta E)}{\rho(E)},
\end{equation}
and we immediately see that \eqref{eq:laplace} ensures that the left-hand side is a properly normalized distribution. The fluctuation-dissipation theorem~\cite{Davis2016} 
associated to $P(\beta|E, S)$ is
\begin{equation}
\label{eq:fdt}
\frac{\partial}{\partial E}\big<\omega\big>_{E,S} = \left<\frac{\partial \omega}{\partial E}\right>_{E,S} + \left<\omega\frac{\partial}{\partial E}\ln P(\beta|E,S)\right>_{E,S}
\end{equation}
which, by replacing \eqref{eq:pbeta_cond}, becomes
\begin{equation}
\label{eq:fdt2}
\frac{\partial}{\partial E}\big<\omega\big>_{E,S} = \left<\frac{\partial \omega}{\partial E}\right>_{E,S} + \Big<\omega(\beta_F-\beta)\Big>_{E,S}
\end{equation}
where we have defined the fundamental inverse temperature function $\beta_F(E)$ by
\begin{equation}
\beta_F(E) \defeq -\frac{\partial}{\partial E}\ln \rho(E).
\end{equation}

Two consequences of the fluctuation-dissipation relation in \eqref{eq:fdt2} are straightforward to obtain. First, by using $\omega = 1$ and recalling that
$\big<f\big>_{E, S} = f(E)$ for any function $f(E)$ of the energy, we immediately see that
\begin{equation}
\label{eq:betaF_super}
\beta_F(E) = \big<\beta\big>_{E,S},
\end{equation}
which then gives meaning to the fundamental inverse temperature in superstatistics: it is the conditional expectation of the superstatistical inverse temperature given
the energy of the system. Second, by taking expectation of \eqref{eq:betaF_super} under $S$ on both sides, we obtain
\begin{equation}
\label{eq:equaltemp}
\big<\beta_F\big>_S = \big<\beta\big>_S,
\end{equation}
that is, the expectation values of $\beta_F$ and $\beta$ coincide, and we can use this common value to define the inverse temperature $\beta_S$ of the ensemble $S$ without 
ambiguity as
\begin{equation}
\label{eq:betaS}
\beta_S \defeq \big<\beta_F\big>_S.
\end{equation}

In the following sections, we will recover the kappa distribution for the single-particle velocity from superstatistics plus just one additional assumption. Furthermore, we 
will show how a superstatistical approximation produces a distribution $P(\beta|S)$ as the thermodynamic limit of the distribution of the inverse fundamental temperature, 
$P(\beta_F|S)$, thus proving a deeper connection between the superstatistical parameter $\beta$ and the function $\beta_F$.

\section{The kappa distribution in steady state plasmas}
\label{sec:kappa}

The total energy of a system of $N$ classical, non-relativistic interacting particles forming a plasma in a steady state can be written as
\begin{equation}
\label{eq:energyfunc}
E(\bm{r}_1, \ldots, \bm{r}_N, \bm{v}_1, \ldots, \bm{v}_N) = \sum_{i=1}^N \frac{m_i \bm{v}_i^2}{2} + \Phi(\bm{r}_1,\ldots,\bm{r}_N),
\end{equation}
in such a way that the details of the interaction with the (self-consistent) electromagnetic fields are contained inside the potential energy
function $\Phi$. This \emph{energy function} $E$ is different from the Hamiltonian $\H$, as the latter should be written in terms of momenta instead of velocities. However, 
in a steady state the joint probability of positions and velocities actually depends only on the energy function $E$ (as we have shown earlier~\cite{Davis2019b}), that is, 
is of the form
\begin{equation}
\label{eq:rho_plasma}
P(\bm{R}, \bm{V}|S) = \rho(E(\bm{R}, \bm{V}); S),
\end{equation}
where we have introduced the shortcut notation $\bm{R} \defeq (\bm{r}_1, \ldots, \bm{r}_N)$ and $\bm{V} \defeq (\bm{v}_1, \ldots, \bm{v}_N)$.
The joint distribution of velocities can be obtained by marginalization of the particle positions,
\begin{equation}
P(\bm{v}_1, \ldots, \bm{v}_N|S) = \int d\bm{R}\,\rho\big(E(\bm{R}, \bm{v}_1, \ldots, \bm{v}_N); S\big) = p_N\Big(\textstyle\sum_{i=1}^N \frac{m_i \bm{v}_i^2}{2}\Big),
\end{equation}
where this relation defines the $N$-particle ensemble function of velocities $p_N$. Moreover, the single-particle velocity distribution, which is our main target in this
work, is given by marginalization in $P(\bm{v}_1, \ldots, \bm{v}_N|S)$ of the remaining $N-1$ particle velocities,
\begin{equation}
\label{eq:isotropic}
P(\bm{v}_1|S) = \int d\bm{v}_2\ldots d\bm{v}_N P(\bm{v}_1, \ldots, \bm{v}_N|S) = p_1\Big(\frac{m_1 \bm{v}_1^2}{2}\Big).
\end{equation}

Here it is important to note that \eqref{eq:rho_plasma} together with the form of the energy function in \eqref{eq:energyfunc} will only lead to isotropic velocity distributions 
because then $P(\bm{v}_1|S)$ depends on $\bm{v}_1$ through its magnitude, according to \eqref{eq:isotropic}. By comparing $P(\bm{v}_1|S)$ with the kappa distribution in 
\eqref{eq:kappa}, we see that our single-particle ensemble function $p_1$ must be given by
\begin{equation}
\label{eq:kappa_p1}
p_1(k_1) = \frac{1}{\eta_\kappa(\vth)}\left[1+\frac{2 k_1}{m_1\vth^2 (\kappa-\frac{3}{2})}\right]^{-(\kappa+1)},
\end{equation}
where $k_1 \defeq m_1\bm{v}_1^2/2$ is the kinetic energy of the particle with $i = 1$. In the next section, we will arrive at the kappa form for $p_1(k_1)$ using a single 
requirement on the dependence between the kinetic energy $k_1$ of a particle and the kinetic energy $K$ of its surrounding environment.

\section{Derivation of the kappa distribution}
\label{sec:kappafromsup}

In the following analysis, we will be considering a group of $n \leq N$ particles as a subsystem, regarding only their kinetic energy. Without loss of generality we can take 
the first particle as a test particle with kinetic energy $k_1$, and the remaining $n-1$ particles as its environment with kinetic energy
\begin{equation}
K \defeq \sum_{i=2}^n \frac{m_i \bm{v}_i^2}{2}.
\end{equation}

Then, the energy $\K$ of the subsystem is directly $\K \defeq k_1 + K$. Recalling that the density of states of kinetic energy for a group of $n$ particles is given by
\begin{equation}
\label{eq:omegan}
\Omega_n(\K) \defeq \int d\bm{v}_1\ldots d\bm{v}_n \delta\Big(\K-\sum_{i=1}^n \frac{m_i\bm{v}_i^2}{2}\Big) = W_n\;\K^{\frac{3n}{2}-1}
\end{equation}
where we have defined the constants
\begin{equation}
W_n \defeq \frac{(2\pi)^{\frac{3n}{2}}M^{-\frac{3}{2}}}{\Gamma\big(\frac{3n}{2}\big)}
\end{equation}
and $M \defeq \prod_{i=1}^n m_i$, the partition function associated to $\Omega_n$ is its Laplace transform,
\begin{equation}
Z_n(\beta; M) = \int_0^\infty d\K \Omega_n(\K)\exp(-\beta \K) = W_n\beta^{-\frac{3n}{2}}\Gamma\Big(\frac{3n}{2}\Big)
= \big(2\pi\big)^{\frac{3n}{2}}M^{-\frac{3}{2}}\beta^{-\frac{3n}{2}},
\end{equation}
which contains the single-particle partition function $Z_1(\beta; m)$ as a particular case with $n = 1$ and $M = m$,
\begin{equation}
\label{eq:zeta1}
Z_1(\beta; m) = \left(\sqrt{\frac{2\pi}{m}}\right)^3\beta^{-\frac{3}{2}}.
\end{equation}

Now we will show that only one condition is sufficient to obtain the kappa distribution for a single particle in a plasma, namely that the
most probable kinetic energy $k^*$ of the test particle given the kinetic energy $K$ of its $(n-1)$-particle environment is linear in $K$.
In more precise terms, we require that
\begin{equation}
\label{eq:requirement}
k^* \defeq \argmax_{k_1} P(k_1|K, S) = \gamma_n + \alpha_n K,
\end{equation}
where the parameters $\gamma_n$ and $\alpha_n$ are functions of $n$. In order to show that \eqref{eq:requirement} leads to the kappa distribution, let us first 
compute the joint distribution $P(k_1, K|S)$ of test particle plus environment, which is given by
\begin{equation}
\begin{split}
P(k_1, K|S) & = \left<\delta\Big(k_1-\frac{m_1\bm{v}_1^2}{2}\Big)\delta\Big(K-\sum_{i=2}^n \frac{m_i\bm{v}_i^2}{2}\Big)\right>_S \\
& = \int d\bm{v}_1\ldots d\bm{v}_n p_n\Big(\sum_{i=1}^n \frac{m_i\bm{v}_i^2}{2}\Big)\delta\Big(k_1-\frac{m_1\bm{v}_1^2}{2}\Big)\delta\Big(K-\sum_{i=2}^n \frac{m_i\bm{v}_i^2}{2}\Big) \\
& = p_n(k_1+K)\Bigg[\int d\bm{v}_1 \delta\Big(k_1-\frac{m_1\bm{v}_1^2}{2}\Big)\Bigg]\left[\int d\bm{v}_2\ldots d\bm{v}_n \delta\Big(K-\sum_{i=2}^n \frac{m_i\bm{v}_i^2}{2}\Big)\right],
\end{split}
\end{equation}
and that by using the definition of $\Omega_n$ in \eqref{eq:omegan}, becomes
\begin{equation}
\label{eq:joint_k1_K}
P(k_1, K|S) = p_n(k_1+K)\Omega_1(k_1)\Omega_{n-1}(K).
\end{equation}

\noindent
The conditional distribution $P(k_1|K, S)$ appearing in \eqref{eq:requirement} can then be obtained as
\begin{equation}
\label{eq:pk1_cond}
P(k_1|K, S) = \frac{P(k_1, K|S)}{P(K|S)} = \frac{p_n(k_1+K)\Omega_1(k_1)}{p_{n-1}(K)}
\end{equation}
where a factor $\Omega_{n-1}(K)$ has been cancelled, and the single-particle density of states $\Omega_1(k_1)$ is readily obtained from \eqref{eq:omegan} with $n = 1$,
\begin{equation}
\label{eq:omega1}
\Omega_1(k_1) = \frac{2}{\sqrt{\pi}}\Big(\frac{2\pi}{m}\Big)^{3/2}\sqrt{k_1}.
\end{equation}

\noindent
Now, because $k^*$ is the argument of the maximum of $P(k_1|K, S)$ according to \eqref{eq:requirement}, it follows that $k^*$ is the solution of the extremum equation
\begin{equation}
0 = \left[\frac{\partial}{\partial k_1}\ln P(k_1|K, S)\right]_{k_1 = k^*},
\end{equation}
and by replacing \eqref{eq:pk1_cond} and \eqref{eq:omega1} we obtain
\begin{equation}
\label{eq:k1_extremum}
\beta_F^{(n)}(k^* + K) = \frac{1}{2k^*},
\end{equation}
where $\beta_F^{(n)}$ is the fundamental inverse temperature of the group of $n$ particles, defined by
\begin{equation}
\beta_F^{(n)}(\mathcal{K}) \defeq -\frac{\partial}{\partial \mathcal{K}}\ln p_n(\mathcal{K}).
\end{equation}

\noindent
We can replace $k^*$ in \eqref{eq:k1_extremum} in terms of $K$ using \eqref{eq:requirement} and, after some algebra, obtain
\begin{equation}
\label{eq:betaFn}
\beta_F^{(n)}(\mathcal{K}) = \frac{\alpha_n+1}{2(\gamma_n + \alpha_n \mathcal{K})},
\end{equation}
from which we can recover the $n$-particle ensemble function $p_n$ by integration,
\begin{equation}
\label{eq:pn}
p_n(\mathcal{K}) = p_n(0)\exp\left(-\frac{\alpha_n + 1}{2}\int_0^\mathcal{K} \frac{d\epsilon}{\gamma_n + \alpha_n \epsilon}\right)
= p_n(0)\left[1 + \Big(\frac{\alpha_n}{\gamma_n}\Big)\mathcal{K}\right]^{-\frac{1}{2\alpha_n}-\frac{1}{2}},
\end{equation}
where $p_n(0)$ is a normalization constant to be determined. By marginalizing $K$ in \eqref{eq:joint_k1_K} and using \eqref{eq:edist_rho} as
\begin{equation}
\label{eq:probk1}
P(k_1|S) = p_1(k_1)\Omega_1(k_1)
\end{equation}
we see that
\begin{equation}
\label{eq:p1_integral}
p_1(k_1) = \int_0^\infty dK p_n(k_1+K)\Omega_{n-1}(K).
\end{equation}

\noindent 
Now, making use of the definite integral
\begin{equation}
\int_0^\infty dy\;y^m \Big[1+r(x+y)\Big]^{-c} = r^{-m-1} B\big(c-m-1, m+1\big)\cdot\Big[1+rx\Big]^{m+1-c}
\end{equation}
for $x > 0$, $r > 0$, $m > -1$ and $c > m+1$ with $B(a, b) \defeq \int_0^1 dt\, t^{a-1}(1-t)^{b-1}$ the Beta function, we finally arrive at
\begin{equation}
\label{eq:p1_incomplete}
p_1(k_1) = p_1(0)\left[1+\Big(\frac{\alpha_n}{\gamma_n}\Big)k_1\right]^{\frac{3n}{2}-\frac{1}{2\alpha_n}-2}.
\end{equation}

By comparing \eqref{eq:p1_incomplete} and \eqref{eq:kappa_p1} we see that we have recovered the kappa distribution for the test particle. However, the dependence of
$\alpha_n$ and $\gamma_n$ with $n$ is not yet known. Because superstatistics imposes, through \eqref{eq:laplace}, that
\begin{equation}
\label{eq:p1_kappa2}
p_1(k_1) = \left(\sqrt{\frac{m}{2\pi}}\right)^3\int_0^\infty d\beta P(\beta|S)\exp(-\beta k_1)\beta^{\frac{3}{2}}
\end{equation}
and we have already shown that $P(\beta|S)$ is size-independent, then $p_1(k_1)$ must also be size-independent, even when $\alpha_n$ and $\gamma_n$ are functions of $n$.
This allows us to define new size-independent parameters $u$ and $\beta_S$ such that
\begin{subequations}
\begin{align}
\frac{1}{u} & \defeq \frac{1}{2} - \frac{3n}{2} + \frac{1}{2\alpha_n}, \\
\beta_S & \defeq \frac{\alpha_n}{u\,\gamma_n},
\end{align}
\end{subequations}
and whose meaning will be revealed shortly. In terms of these parameters, we can rewrite \eqref{eq:p1_incomplete} as
\begin{equation}
\label{eq:p1}
p_1(k_1) = p_1(0)\Big[1+(u\beta_S)k_1\Big]^{-(\frac{1}{u}+\frac{3}{2})}.
\end{equation}

Comparison with \eqref{eq:kappa_p1} gives the usual parameters $\kappa$ and $\vth$ of the kappa distribution for a single particle in terms of $u$ and $\beta_S$ as
\begin{subequations}
\begin{align}
\kappa & = \frac{1}{u}+\frac{1}{2}, \\
\label{eq:vth_new}
\frac{m\vth^2}{2} & = \frac{1}{(1-u)\beta_S},
\end{align}
\end{subequations}
and we can use these new parameters $u$ and $\beta_S$ to rewrite the fundamental inverse temperature $\beta_F^{(n)}(\mathcal{K})$ in \eqref{eq:betaFn} as
\begin{equation}
\label{eq:betaFn_new}
\beta_F^{(n)}(\mathcal{K}) = \Big(1+\frac{3nu}{2}\Big)\left[\frac{\beta_S}{1+u\beta_S\mathcal{K}}\right].
\end{equation}
 
We see that $u \rightarrow 0$, that is, $\kappa \rightarrow \infty$, reduces $\beta_F^{(n)}(\mathcal{K})$ to the constant function equal to $\beta_S$ for all $\K$, thus 
recovering the canonical ensemble. Replacing \eqref{eq:p1} and \eqref{eq:omega1} into \eqref{eq:probk1} we obtain the single-particle energy distribution, which after 
normalization yields
\begin{equation}
\label{eq:probk1_explicit}
P(k_1|u, \beta_S) = \frac{\big(\sqrt{u\beta_S}\big)^3}{B\Big(\frac{3}{2}, \frac{1}{u}\Big)}\Big[1+(u\beta_S)k_1\Big]^{-(\frac{1}{u}+\frac{3}{2})}\sqrt{k_1},
\end{equation}
result that fixes the normalization constant $p_1(0)$ to be
\begin{equation}
\label{eq:p1_0}
p_1(0) = \left(\sqrt{\frac{m u\beta_S}{2\pi}}\right)^3\frac{\Gamma(\frac{1}{u}+\frac{3}{2})}{\Gamma(\frac{1}{u})},
\end{equation}
in full agreement with $p_1(0) = \eta_\kappa^{-1}$ as it appears in \eqref{eq:etakappa}. The mean and relative variance of $P(k_1|u, \beta_S)$ in \eqref{eq:probk1_explicit} 
are given by
\begin{subequations}
\begin{align}
\label{eq:mean_k1}
\big<k_1\big>_{u,\beta_S} & = \frac{3}{2\beta_S(1-u)}, \\
\label{eq:var_k1_rel}
\frac{\big<(\delta k_1)^2\big>_{u,\beta_S}}{\big<k_1\big>_{u,\beta_S}^2} & = \frac{2+u}{3(1-2u)},
\end{align}
\end{subequations}
and from these two equations we can, in principle, determine $u$ and $\beta_S$ from the observed statistics of $k_1$. Note that the relative variance in \eqref{eq:var_k1_rel} 
increases monotonically with $u$ from its value of 2/3 for $u = 0$. Additionally, we see that in order to keep $\big<(\delta k_1)^2\big>_{u,\beta_S}$ a non-negative quantity, 
it is required that $u < 1/2$, that is, the spectral index $\kappa$ must be larger than 5/2. Again, in the limit $u \rightarrow 0$ we can confirm, using 
\[\lim_{u \rightarrow 0} \Big[1+(u\beta_S)k_1\Big]^{-(\frac{1}{u}+\frac{3}{2})} = \exp(-\beta_S k_1),\] that $P(k_1|u, \beta_S)$ in \eqref{eq:probk1_explicit} reduces 
to the Maxwell-Boltzmann distribution of single-particle energies,
\begin{equation}
\label{eq:MBE}
P(k_1|\beta) = \left(\frac{2}{\sqrt{\pi}}\right)\beta^{\frac{3}{2}}\exp(-\beta k_1)\sqrt{k_1}
\end{equation}
with $\beta = \beta_S$. Similarly, using \eqref{eq:edist_rho} as $P(\mathcal{K}|u,\beta_S, n) = p_n(\mathcal{K})\Omega_n(\mathcal{K})$ we obtain the energy distribution 
for the group of $n$ particles as
\begin{equation}
\label{eq:pK}
P(\mathcal{K}|u, \beta_S, n) = \frac{\big(\sqrt{u\beta_S}\big)^{3n}}{B\Big(\frac{3n}{2}, \frac{1}{u}\Big)}\Big[1+u\beta_S \mathcal{K}\Big]^{-\big(\frac{1}{u}+\frac{3n}{2}\big)} \mathcal{K}^{\frac{3n}{2}-1},
\end{equation}
and we can verify that
\begin{equation}
\big<\K\big>_{u,\beta_S} = \frac{3n}{2\beta_S(1-u)} = n\big<k_1\big>_{u,\beta_S},
\end{equation}
hence the mean kinetic energy is an extensive quantity for all $n > 1$ and for all $u$. By simple inspection we can also confirm that \eqref{eq:pK} includes 
\eqref{eq:probk1_explicit} as a particular case with $n = 1$ and $\mathcal{K} \rightarrow k_1$. 

We can gain further insight on the relationship between $k^*$ and $K$ if we write our original requirement in \eqref{eq:requirement} in terms of $u$, $\beta_S$ and $n$ as
\begin{equation}
k^*(K) = \frac{1+u\beta_S K}{\beta_S\big([3n-1]u +2\big)}.
\end{equation}

We readily see that the only case where $k^*$ is independent of $K$ corresponds to $u = 0$, that is, to the canonical ensemble with
\begin{equation}
\beta_S = \frac{1}{2 k^*},
\end{equation}
while for $u > 0$ in the thermodynamic limit we have
\begin{equation}
\lim_{n \rightarrow \infty} k^*(K) = \lim_{n \rightarrow \infty} \frac{K}{3(n-1)} = \frac{k}{3},
\end{equation}
where we have defined $k \defeq \lim_{n \rightarrow \infty} K/(n-1)$ as the average kinetic energy of the environment. This is in agreement with the 
mode and mean of the Maxwell-Boltzmann distribution of energies in \eqref{eq:MBE}, namely
\begin{equation}
k^*(\beta) = \frac{1}{2\beta} = \frac{1}{3}\big<k_1\big>_\beta.
\end{equation}

\noindent
On the other hand, the joint distribution $P(k_1, K|u, \beta_S)$ in \eqref{eq:joint_k1_K} yields the covariance between $k_1$ and $K$ as
\begin{equation}
\big<\delta k_1 \delta K\big>_{u,\beta_S} = \frac{9u(n-1)}{4\beta_S^2 (1-u)^2 (1-2u)} \geq 0,
\end{equation}
with equality only for $u = 0$. We can check that this covariance increases monotonically with $u$, and that $k_1$ and $K$ are statistically independent if and only 
if $u = 0$.

\section{Statistical distribution of inverse temperatures}

The superstatistical distribution of the inverse temperature $\beta$, namely $P(\beta|u, \beta_S)$, can now be determined by using \eqref{eq:supweight} in the form
\begin{equation}
P(\beta|u, \beta_S) = f_1(\beta)Z_1(\beta)
\end{equation}
with $f_1 = \mathcal{L}^{-1}\{p_1\}$ the inverse Laplace transform of the single-particle ensemble function $p_1$ in \eqref{eq:p1}. Because the
inverse Laplace transform is unique if it exists, and recalling the Euler integral
\begin{equation}
\int_0^\infty d\beta \exp(-\beta A)\beta^{R-1} = \Gamma(R)A^{-R},
\end{equation}
we obtain for $A = k_1 + 1/(u\beta_S)$ and $R = 1/u + 3/2$, that
\begin{equation}
f_1(\beta) = \frac{p_1(0)}{u\beta_S\Gamma(\frac{3}{2}+\frac{1}{u})}\exp\left(-\frac{\beta}{u\beta_S}\right)\left(\frac{\beta}{u\beta_S}\right)^{\frac{1}{u}+\frac{1}{2}}.
\end{equation}

After multiplying by $Z_1(\beta)$ in \eqref{eq:zeta1} and replacing \eqref{eq:p1_0}, we obtain the properly normalized probability distribution for $\beta$ as
\begin{equation}
\label{eq:pbeta}
P(\beta|u, \beta_S) = \frac{1}{u\beta_S\;\Gamma(1/u)}\exp\left(-\frac{\beta}{u\beta_S}\right)\left(\frac{\beta}{u\beta_S}\right)^{\frac{1}{u}-1},
\end{equation}
which is a gamma distribution with mean and variance given by
\begin{subequations}
\begin{align}
\big<\beta\big>_{u,\beta_S} & = \beta_S, \\
\big<(\delta \beta)^2\big>_{u, \beta_S} & = u(\beta_S)^2.
\end{align}
\end{subequations}

Here we see that $\beta_S$ is directly the mean superstatistical inverse temperature, in agreement with \eqref{eq:betaS} and \eqref{eq:equaltemp}, while $u$ is the relative 
variance of $\beta$, thus together with $u < 1/2$ we see that we must have $0 \leq u < 1/2$. The most probable inverse temperature is given by
\begin{equation}
\label{eq:betaS_star}
\beta_S^* \defeq \beta_S(1-u),
\end{equation}
and it is clear that $u \rightarrow 0$ recovers the canonical ensemble, because
\begin{align}
\big<(\delta \beta)^2\big>_{u,\beta_S} & \rightarrow 0, \\
\beta^*_S & \rightarrow \beta_S,
\end{align}
which together imply $P(\beta|u, \beta_S) \rightarrow \delta(\beta-\beta_S)$, in agreement with the limit $\kappa \rightarrow \infty$ of the kappa distribution, i.e. the Maxwell-Boltzmann 
distribution. Furthermore, using \eqref{eq:betaS_star} and letting $k_B T^*_S \defeq 1/\beta^*_S$, we can rewrite \eqref{eq:vth_new} as
\begin{equation}
\vth = \sqrt{\frac{2k_B T^*_S}{m}},
\end{equation}
which agrees with \eqref{eq:vth} if we interpret the parameter $T$ appearing in the kappa distribution as $T^*_S$ of the superstatistical description. The conditional distribution 
of inverse temperature given $K$ follows from Bayes' theorem as
\begin{equation}
P(\beta|K, u, \beta_S, n) = \frac{P(\beta|u,\beta_S)P(K|\beta)}{P(K|u,\beta_S,n)} = \frac{P(\beta|u, \beta_S)\exp(-\beta K)}{Z_{n-1}(\beta)\,p_{n-1}(K)},
\end{equation}
where we have cancelled a factor $\Omega_{n-1}(K)$. This is also a gamma distribution, written explicitly as
\begin{equation}
P(\beta|K, u, \beta_S, n) = \frac{\Big[1 + u\beta_S K\Big]^{\frac{1}{u}+\frac{3(n-1)}{2}}}{u\beta_S\Gamma\big(\frac{1}{u}+\frac{3(n-1)}{2}\big)}\exp\left(-\frac{\beta}{u\beta_S}
\Big[1 + u\beta_S K\Big]\right)\left(\frac{\beta}{u\beta_S}\right)^{\frac{1}{u}+\frac{3(n-1)}{2}-1},
\end{equation}
but, unlike $P(\beta|u, \beta_S)$ in \eqref{eq:pbeta}, this distribution is explicitly dependent on the size $n$. The mean inverse temperature given $K$ is
\begin{equation}
\big<\beta\big>_{K, u, \beta_S, n} = \left(1 + \frac{3(n-1)u}{2}\right)\left[\frac{\beta_S}{1+u\beta_S K}\right],
\end{equation}
and, by comparing with \eqref{eq:betaFn_new}, we can verify that \eqref{eq:betaF_super} holds, in the form
\begin{equation}
\big<\beta\big>_{K, u, \beta_S, n} = \beta_F^{(n-1)}(K).
\end{equation}
 
This means $\big<\beta\big>_{K, u,\beta_S, n}$ also reduces to $\beta_S$ in the limit $u \rightarrow 0$ with finite $n$, becoming independent of $K$. In the 
thermodynamic limit, that is, when $n \rightarrow \infty$, we have that
\begin{equation}
\label{eq:lim_betaKS}
\lim_{n \rightarrow \infty} \big<\beta\big>_{K, u, \beta_S, n} = \lim_{n \rightarrow \infty} \frac{3(n-1)}{2K} = \frac{3}{2k}
\end{equation}
for $u > 0$. The relative variance of $P(\beta|K, u, \beta_S, n)$ is
\begin{equation}
\frac{\big<(\delta \beta)^2\big>_{K, u, \beta_S, n}}{\big<\beta\big>_{K, u, \beta_S, n}^2} = \frac{2u}{2+3(n-1)u},
\end{equation}
and vanishes both in the limit $u \rightarrow 0$ and in the thermodynamic limit with $u > 0$, unlike the relative variance of $P(\beta|u, \beta_S)$ which is 
independent of $n$. This last result, combined with \eqref{eq:lim_betaKS}, implies that
\begin{equation}
\label{eq:limit_delta}
\lim_{n \rightarrow \infty} P(\beta|K, u, \beta_S, n) = \delta\Big(\beta - \frac{3}{2k}\Big).
\end{equation}

We can interpret this result as the following statement: in the thermodynamic limit, the kinetic energy of a group of particles uniquely fixes its superstatistical temperature, 
and this temperature becomes exactly the fundamental temperature. 

\section{Summary and discussion}
\label{sec:concluding}

We have shown that the kappa distribution for particle velocities in a plasma can be recovered from superstatistics plus a single assumption, namely 
\eqref{eq:requirement} which imposes linearity of the most probable kinetic energy $k^*$ of a test particle as a function of the kinetic energy $K$ of its 
environment. Our results do not rely on the concept of entropy or its maximization, non-additivity or any such concept, and do not assume any particular distribution 
of temperature \emph{a priori}. Nevertheless, in such a plasma the inverse temperature $\beta$ does have a well-defined distribution, namely the gamma distribution $P(\beta|u, \beta_S)$ in \eqref{eq:pbeta}. 

Our result shows that the kappa distribution can arise whenever there are kinetic energy correlations, suggesting that it may be realized in more diverse experimental conditions than are currently considered. Relevant new scenarios to be explored may include laser-produced plasmas~\cite{Harilal2022}, Z-pinches~\cite{Haines2011} and in particular plasma focus
devices~\cite{Soto2005, Soto2010, Auluck2021}, where a rich phenomenology has been observed, including dense plasma~\cite{Tarifeno2010}, plasma shocks~\cite{Soto2014}, plasma filaments~\cite{Pavez2022} and supersonic plasma jets~\cite{Soto2005,Soto2010,Auluck2021,Tarifeno2010}. The recently postulated relationship between 
the mechanism of magnetic reconnection and kappa distributions~\cite{Hoshino2022, Hoshino2023} suggests that this distribution may also describe the emission 
of plasma foci, as magnetic reconnection may also be a relevant process in those devices~\cite{Kubes2020}.

An open question, left for future studies, is the possibility that a similar mechanism of constraining the correlations between observables may lead to the 
other two universality classes in superstatistics, namely log-normal and inverse gamma forms for $f(\beta)$.

\section*{Acknowledgements}

\noindent
S.D. gratefully acknowledges funding from ANID FONDECYT 1220651 grant.


\end{document}